\documentclass[reprint, amsmath,amssymb,aps,longbibliography,nofootinbib]{revtex4-1}

\usepackage[english]{babel}
\usepackage[autostyle, english = british]{csquotes}
\usepackage[CJKbookmarks, pdftex, bookmarksnumbered, bookmarksopen, colorlinks, citecolor=blue, linkcolor=blue]{hyperref}
\usepackage{amsmath,amssymb}
\usepackage{graphicx}
\usepackage{enumitem}
\usepackage[capitalize]{cleveref}
\setlist[enumerate]{topsep=0pt,parsep=-1mm,leftmargin=5mm,}
\def\be{\begin{equation}}
\def\ee{\end{equation}}

\usepackage{color} 
\RequirePackage[dvipsnames,usenames]{xcolor}

\newcommand{\eq}[1]{\begin{align}#1\end{align}}

\newcommand{\D}{{\cal{D}}}
\newcommand{\XX}{{\cal{X}}}
\newcommand{\E}{{\cal{E}}}
\newcommand{\SSS}{{\cal{S}}}
\newcommand{\TT}{{\cal{T}}}

\newcommand{\PD}{{\cal{D}}}

\begin{document}

\title{\large Disentangling Boltzmann brains, the time-asymmetry of memory, and the second law}

\author{David Wolpert${}^{a}$, Carlo Rovelli${}^{abcd}$,  Jordan Scharnhorst${}^{e}$}
\affiliation{${}^{a}$Santa Fe Institute, 1399 Hyde Park Road Santa Fe, New Mexico 87501, USA} 
\affiliation{${}^{b}$Aix-Marseille University, Universit\'e de Toulon, CPT-CNRS, F-13288 Marseille, France.}
\affiliation{${}^{c}$Department of Philosophy and the Rotman Institute of Philosophy, 1151 Richmond St.~N London N6A5B7, Canada}
\affiliation{${}^{d}$Perimeter Institute, 31 Caroline Street N, Waterloo ON, N2L2Y5, Canada} 
\affiliation{${}^{e}$University of California, Santa Cruz, 1156 High St., Santa Cruz, California, 95060}

\begin{abstract} 
\noindent 
Are your perceptions, memories and observational data, a
statistical fluctuation out of the thermal equilibrium of the universe, having no correlation
with the actual past state of the universe? Arguments are given in the literature 
for and against this ``Boltzmann brain'' hypothesis. Complicating these arguments have been the many subtle -- and very often
implicit -- joint dependencies among these arguments and others that have been given for the past hypothesis,
the second law, and even for Bayesian inference of the reliability of experimental data. 
These dependencies can easily lead to circular reasoning.
To avoid this problem, since all of these arguments involve the stochastic properties of the
dynamics of the universe's entropy, we begin by formalizing that dynamics
as a time-symmetric, time-translation invariant Markov process, which we call
the \textbf{entropy conjecture}.
Crucially, like all stochastic processes, the entropy conjecture
does not specify any
time(s) which it should be conditioned on in order to infer the stochastic dynamics of our universe’s entropy. 
Any such choice of conditioning times and associated entropy values must be introduced as an independent assumption. This observation
allows us to disentangle the standard Boltzmann brain hypothesis, its ``1000CE" variant,
the past hypothesis, the second law, and the reliability of our experimental data, all
in a fully formal manner. In particular, we show that these all 
make an arbitrary assumption that the dynamics of the universe's entropy
should be conditioned
on a single event at a single moment in time, differing only in the details of their assumptions. 
In this aspect, the Boltzmann brain hypothesis and the second law 
are equally legitimate (or not).
\end{abstract}

\maketitle

{\section{Introduction}}
\noindent 

Consider a large statistical system formed by a mixture of particles of different kinds that remains in thermal equilibrium for an arbitrarily long span of time.  According to statistical mechanics, there are fluctuations at thermal equilibrium, and in principle all configurations can be reached by such fluctuations with enough time available. Consider one of these random fluctuations giving rise -- just by chance -- precisely to a brain like ours, complete with our memories, information and perception, but no corresponding external physical world. 
Such a brain -- indistinguishable from ours, but arising as a fleeting fluctuation -- is call a ``Boltzmann brain'' (BB) \cite{albrecht2004can,linde2007sinks,aguirre2012out}). Such a brain would have exactly our current memories, perceptions, and it would know and feel precisely what we know and feel right now. 

Is it possible to know that we are not a BB?
An intuitive answer is that we yes, we can know this, that we are not a BB,
since the probability for such a fluctuation to occur is fantastically small, and the expected time for this to happen is colossally longer than the current age of the universe. However, recent literature \cite{albrecht2004can,linde2007sinks,GoheerTrouble,DysonImplications,PageReturn} has questioned this simple answer. Counter-counter arguments have then said that the reasoning behind the BB hypothesis is ``cognitively unstable'', 
that in fact it is ``unstable reasoning''\cite{myrvold2016probabilities,carroll.simulation.hypothesis,elga2025boltzmann,wallace2023bayesian}.


To understand the issues involved in these disputes, it is worth describing
the standard argument for the BB hypothesis in a bit more detail. 
To simplify matters, for the most part we restrict attention to classical physics, since much
of the controversy also imposes that restriction. The associated standard argument for the BB hypothesis
starts by pointing out that we currently have two  particular sets of information:
\begin{enumerate}[nosep, label=(\roman*)]
\item A particular set $\cal L$ of microscopic laws of physics, including in particular Newton's laws. Note that all of these 
laws are symmetric under time inversion.
\item The set $\cal D$ of all present observations and that we have concerning the physical
universe. This includes all of our current records, recorded data, and memories.
\end{enumerate} 
What can we say about the physical universe, based on these two sets of information? 
The set of laws $\cal L$ admits a large family of solutions.  For simplicity, for the moment we  
restrict ourselves to these solutions that are consistent with $\cal D$, in the sense that $\cal D$ has non-infinitesimal likelihood
under those solutions. Similarly, let us assign equal probability to all these solutions that we restrict ourselves to. 

Building on the widely accepted set of laws of modern classical and quantum physics, 
the fluctuation theorems of stochastic thermodynamics and open quantum
thermodynamics have established that the evolution of entropy in all statistical physics systems,
is a stochastic process~\cite{shiraishi2023introduction,peliti2021stochastic,seifert2025stochastic}.
It is also widely accepted that (in non-cosmological contexts) this process:
\begin{enumerate}
\item is time translation invariant;
\item is symmetric under time-inversion;
\item has the property that if it is conditioned on a value of entropy $s_0$ at one
particular time, $t_0$, and not conditioned on anything else, and if  $s_0$ is sufficiently lower than the maximal
value of entropy, then the expected value of entropy for times greater than $t_0$ are greater than $s_0$.
\end{enumerate}
We call this set of three presumptions the \textbf{entropy conjecture}. We call any process
satisfying the first two of these presumptions a \textbf{Boltzmann process} (whether or not it involves physical entropy).


In addition to implying that the universe's entropy evolves as a Boltzmann process, $\cal D$ includes
the current value of the universe's entropy, and in particular the fact that it is far lower than its maximum.
(In other words, our current data tells us the current value of the universe's entropy, albeit only
with limited precision.)
Crucially, the H  theorem says that with extraordinarily high probability, entropy increases both into the future \textit{and into the past},
from any such time at which entropy is known to have some specific low value. (Indeed,
the entropy conjecture is time-symmetric about such a special time.) In other words,
the most probable situation, given the current observations, is that we happen to be precisely at a special 
point in the dynamics of the universe's entropy. In other words, the most probable situation is that we are just an entropy fluctuation, which is to say that we are a BB. 

This standard argument for the BB hypothesis initially arose in the context of de Sitter quantum cosmology \cite{albrecht2004can,linde2007sinks,GoheerTrouble,DysonImplications}. If such a universe evolves into the future, it asymptotes to a vacuum state with a finite, non-zero temperature. This equilibrium state has fluctuations that act similarly to thermal fluctuations in a classical gas at equilibrium, which satisfy an entropy conjecture and corresponding recurrence theorems. 

Indeed, suppose that the set of laws $\cal L$ imply thermalization of the universe in the distant future, so that
 the universe settles into a long living equilibrium state (Boltzmann's ``thermal death")
Then even if the universe started off in a low entropy state, $\cal D$ will be realized an infinite number of times (up to arbitrary precision) in the distant future. This then raises the question of how we could conclude that with our data $\cal D$ are not one of these fluctuations, 
which occur with probability $1$. 

One may respond to these arguments that we also know the second law of thermodynamics, and from this we know that entropy 
must grow with $t$. This directly contradicts the BB hypothesis.
%
However, our knowledge of the second law arose from consideration of our data records about the past: how do we know that these are data records of the past and aren't {\em themselves} due to a fluctuation?  
A common response is to simply assume that entropy was in fact low in the past and has been growing since.  But this  amounts to a (very) improbable assumption, given the entropy conjecture and $\cal D$. Those data $\cal D$ are vastly
more  likely to be the result of a fluctuation than to be the result of low entropy in the distant past.  

Round and round the arguments go. A common problem with all these arguments is that they are, 
at best semi-formal. The only way to resolve their apparent conflicts, is to use fully rigorous
reasoning, grounded in stochastic process theory. That is what we do here.

In a nutshell, our analysis starts by recognizing that what Boltzmann
proved (though he did not have the tools to formulate this way)
is that the stochastic process of entropy (or in his formulation, the H
quantity) is a time-stationary Markov process, symmetric under
time inversion, which is biased towards high values. 

Crucially, to compute a distribution over possible trajectories of a random variable
that evolves according to \textit{any} stochastic process, it is first
necessary to say precisely what (if anything) that distribution is conditioned on.
In the context of the entropy conjecture, that means specifying a (perhaps
empty) set of pairs of a value of entropy, and
an associated time when the universe had that value of entropy. 

Any such specified conditioning events can\textit{not} arise
from the entropy conjecture. In fact, it is not provided by physics. Rather 
such events must be determined using Bayesian 
reasoning, based on our current observational data and the prior
we adopt over the possible laws of the universe. 

As we show, this simple observation, formulated in a fully
rigorous manner, disentangles all the ambiguities and circular
reasoning that are rife in the literature on Boltzmann brains, and
more generally in the literature on the past hypothesis (PH), and even in
the literature on the second law. In particular, our analysis shows
that the PH and the BB hypothesis are, formally speaking identical,
in that both only condition on a subset of our current data,
differing only in which such data they condition on. We also
show that arguments about ``unstable reasoning'' do not in fact prove
anything concerning the validity of the BB hypothesis, only about one
particular argument that has been offered for that validity.
We also formalize some of the properties of all circular reasoning,
illustrating that they apply to common arguments for the
second law.

We emphasize that we do \textit{not} make any arguments for or
against the BB hypothesis, the PH, or the second law. Indeed, as we
prove, ultimately any such arguments must rely on choices about what
data to condition the H theory on, and that choice cannot be provided
by physics. Rather we
provide a novel, fully formal framework for investigating those concepts
and their very subtle relations.

We also emphasize that we do not dispute the major role of cosmology in the second law of thermodynamics \cite{CarloWhere,davies1977physics,scharnhorst2023} or investigate precisely which models predict BB and which do not. Indeed, there are even major arguments that a framework for studying BB depending strictly on the entropy conjecture, without cosmology, is incomplete. It has long been known that the BB issue is highly sensitive to schema for assigning probabilities to values of the scale factor in FLRW scalar field cosmology models \cite{PageReturn,BrainsCutoffMeasure,VilenkinFreak}. We simply start from the working basis of pure statistical physics to unify and contextualize a recent genre of arguments in the literature.

We begin in~\cref{sec:II} by summarizing one of the more nuanced,
semi-formal counter-arguments that can be made to the BB hypothesis.
A central concern with the counter-argument presented there is how we
can reason from current data to conclusions about how experiments were initialized
in the past, and from there to infer laws of the universe. The key conundrum
we highlight is that
this inference process itself relies on those laws being inferred, and so is circular.
(Arguments in the literature that reach similar conclusions have characterized the standard argument
for the BB hypothesis as ``unstable reasoning''. See~\cite{myrvold2016probabilities,carroll.simulation.hypothesis} for some work related to this counter-argument.)

In the following section we begin by introducing the mathematical machinery 
we will need to disentangle the arguments concerning the BB hypothesis. Our first contribution is to introduce a restriction on sigma algebras that is
necessary if (as in all analyses of of the PH, the second law, etc.) 
we use to investigate probability measures over possible
``laws of physics''. We then show how to fully formalize the entropy conjecture
as a Markov process; this is our second contribution. It builds crucially
on the results in~\cite{scharnhorst2024boltzmann}.

Our third
contribution is to use this formalization of the entropy conjecture to clarify what it cannot tell us. In contrast,
our fourth contribution  is to clarify what the entropy conjecture does in fact
tell us concerning the BB hypothesis specifically. Next, our fifth
contribution is to use our formalization of the entropy conjecture to formalize
the reasoning in~\cref{sec:II}, and to analyze
what arguments concerning unstable reasoning can and cannot tell
us about the BB hypothesis.  After this we present our sixth contribution,
which is a variant of the BB hypothesis we call the ``1000 CE hypothesis''.
As we show, formally speaking,  the 1000 CE hypothesis lies exactly
between the standard BB hypothesis and the PH. This illustrates the
underlying identity between the standard BB hypothesis and the PH.

We end with a discussion, and then present some of the myriad future
areas of research that our investigation suggests.

\section{An objection to the standard argument for Boltzmann Brains}
\label{sec:II}



Recall from above that we write $\cal L$ for the (classical) laws of physics as we currently understand them, and $\cal D$ for our
current observational data. In addition, write $\cal B$ for the Boltzmann brain hypothesis. In addition,
as shorthand we write all statements that a particular conditional probability $P(A \;|\; B)$
is close to $1$ as $A \rightarrow B$.

To recap from the introduction,
the conditional probability $P({\cal B}\,|\,{\cal L},{\cal D})$ of the BB hypothesis, given the laws of physics and the data observed in the present, is near $1$. So in terms of our shorthand, that standard argument says that
\be
({\cal D},{\cal L})\to {\cal B}.
\ee
%
The problem with the standard argument that we wish to highlight is that 
we \textit{infer} $\cal L$ \textit{from} $\cal D$ --- but that inference in turn invokes those very
laws $\cal L$ we wish to infer. Specifically, it is now understood that our brains must rely on the second law of thermodynamics, which is contained
in $\cal L$, to infer that our current data
$\cal D$ accurately records the results of past experimental tests~\cite{wolpert2024memory,wolpert1992memory}.
So in particular, we need to use the second law to infer from our current data the results of our past experiments
of the the second law itself, $\cal L$. So we have circular
reasoning in the very conditioning events we wish to use to ground the standard argument for the BB hypothesis.

To elaborate this problem, 
let's focus on the laws $\cal L$.  How do we know the laws of the universe that we claim to know? By knowing
only some present data $\cal D$, i.e., properties of the current universe, we cannot infer these laws. 
This is even true when $\cal D$ contains the results of experimental tests of those laws.
This is because the 
laws $\cal L$ are relations between the values of the physical variables \emph{at different times}, hence we cannot know or infer anything about them unless we also know data about the world at a time different that the present. In other words,
we also need to know how those experimental tests were initialized in the past,
in addition to the results of those tests, recorded in $\cal D$.
%
%
That is, we don't directly have ${\cal D} \to {\cal L}$, but rather 
\be
({\cal P},{\cal D})\to {\cal L}. 
\label{eq:2}
\ee

How do we access the past data $\cal P$? As described in~\cite{wolpert2024memory,rovelli2014we},
to do this we have to trust the reliability of \emph{records} or \emph{memories} we have at the present that concern these past data. 
For notational simplicity, we indicate these
memories as parts of $\cal D$.
Let us call $\cal R$ the assumption that such memories of past data are statistically reliable,
that they accurately reflect the ways that our experimental apparatus(es) were established
in the past.\footnote{In the terminology of~\cite{wolpert2024memory},
the results of almost all science experiments are recorded in a ``type-3'' memory. The second law can be
applied to deduce information about the past state of the universe from the current state of such a memory, i.e., such memories
are ``reliable''. Accordingly, we can formalize 
$\cal R$ as the statement that we have a type-3 memory, whose current state is specified in $\cal D$,
and that provides the information $\cal P$ about the past state of the universe. } 
So the combination $(\cal{D}, \cal{R})$ establishes $\cal P$.
Therefore writing $(\cal{D}, \cal{R})$ is equivalent to writing $(\cal{D}, \cal{R}, \cal{P})$.

Now, let us consider the two cases separately: whether we make the assumption $\cal R$ or we do not.
%
%
If we do, then by \cref{eq:2} we have 
\be
({\PD} ,{\cal R})\to {\cal L},
\label{questa}
\ee
and therefore $P(\neg {\cal L} \;|\; \PD, \cal R)$ is close to $0$.
Bayes theorem then says
\be
P({\cal B} \,|\,\PD,{\cal R}) \simeq P({\cal B}\,|\,\PD,{\cal R},{\cal L})P({\cal L}\,|\,\PD,{\cal R}),
\ee
Even though the second term on the RHS, $P({\cal L}\,|\,\PD,{\cal R})$, is close to $1$ (by hypothesis), the first term on the RHS,
$P({\cal B}\,|\,\PD,{\cal R},{\cal L})$, is close to $0$. Therefore $P({\cal B}\,|\,\PD,{\cal R})$ is close to $0$.
Therefore, returning to our simplified notation, 
\be
(\PD,{\cal R})\to ({\cal L},\PD,{\cal R})\to ({\rm not}\ {\cal B}).
\ee

What about if we instead do \emph{not} assume $\cal R$?  In this case, we have no argument in support of $\cal L$. Of course $\cal B$ does not follow from $\cal D$ alone, without $\cal L$. 
Without an assumption about reliability of inferences about the past from current data,
i.e., without an assumption about the reliability of memory systems~\cite{wolpert2024memory,rovelli2022memory}, we have no reasons to believe in BBs. Data at a single moment of time are compatible with any dynamical law, namely with both $\cal{B}$ and $\neg \cal{B}$. Absent other assumptions, they characterize the state of the universe at a single time, at best. They are not informative about the laws of the universe. 


One may object: we can consider $\cal L$ not as a component of our knowledge, but rather as a fact of the world. That is, suppose we say: the world is truly governed by the laws $\cal L$, irrespectively from our knowledge.  What can we deduce from this fact combined with the fact that we know $\cal D$? This appears to circumvent the observation that our knowledge of $\cal L$ depends on our knowledge of records and on the assumption $\cal R$. But this objection does not hold. 
By themselves, events either happen or don't. Facts are either true or false. 
To reason about likelihoods can only be based on incomplete knowledge and assumptions. The only reasonable question here is the likelihood that {\em we} assign to the possibility of being BBs, and this only depends on what {\em we} consider reliable. Our considering $\cal L$  to be reliable is grounded in our considering records reliable, hence our not being in a fluctuation.  That is, it necessitates the assumption $\cal R$.
So we do need to assume $\cal R$ after all.

\section{Relation between the entropy conjecture, Boltzmann Brains, and the second law}
\label{sec:III}

The discussion in the preceding sections 
may leave a sense of confusion. Many parts of the arguments recounted above
involving the BB hypothesis might seem to involve circular reasoning. Such reasoning can only (!) establish that the
priors it relies on are inconsistent. However, the fact that an argument engages
in circular reasoning doesn't tell us anything about how valid its \textit{conclusions} are,
just that the priors it relies on don't themselves establish those conclusions.
(See~\cref{sec:app_B} for a Bayesian definition ``circular reasoning'',
and an analysis of what it does and does not imply.)

Another possible concern is that the arguments are only semi-formal, relying on ``common experience''. 
However, in any investigation concerning foundational issues in thermodynamics,
(like whether BBs accord with our present data records),
if one does not use fully formal mathematical reasoning it is \textit{extremely} easy to
subtly assume what one is in fact trying to prove.
For example, that mistake is made in very many of the 
papers in the literature trying to prove the second law. Specifically, the analyses in
many of these papers implicitly privilege the
initial condition of a physical system (e.g., how it was set up) over its final condition (e.g., how it finishes). 
That asymmetry in boundary conditions assumed by those arguments is, ultimately, the basis
for the ``derivation'' of the second law. However, our privileging the initial conditions
in turn arises from the psychological arrow of time -- which it is commonly believed
to be a consequence of the asymmetry of the second law. So such a supposed derivation involves circular reasoning.
The same mistake of circular reasoning has also been made in many earlier investigations
of the BB hypothesis. 

Fortunately, there is a formally rigorous
framework that was recently introduced in the context of proving the second law 
specifically to avoid this mistake of circular reasoning~\cite{scharnhorst2024boltzmann}.
In the rest of this section we build on that rigorous framework, to
fully formalize our arguments concerning the second law and BBs in earlier sections. 
This formal rigor ensures that there are no ``gotcha's''
hiding in our central argument, and in particular that we do not implicitly assume 
the very thing we are trying to prove.
This formalization also makes clear just how different our argument is from any 
that has previously been considered in the literature. 

Before presenting this formal version of our central argument,
we need to say something about our terminology. 
Throughout this text we will often refer to ``Newtonian laws'' as shorthand for all of classical physics.
With some more care about the precise meaning of ``space-time events'', this shorthand could be
extended even further to also include quantum mechanics and general relativity.

\subsection{Using probability distributions to reason about Newton's laws}

To begin, note that much of the literature (and in particular our central argument)
ascribes probabilities to various possible laws of physics, e.g., to the second law.
However, to meaningfully speak about a {probability distribution}
over {{laws}}, we need to at least sketch how physical laws can be formulated in terms of 
sets in a sigma algebra that that probability distribution is defined over. 
In particular, to fully clarify the (in)validity of the BB hypothesis, the PH, and the second law,
we need to pay scrupulous attention to how we reason from current data to 
probabilistic conclusions concerning the 
laws of the universe.


In this paper, for completeness, we provide a rigorous definition of what it means to have a probability
measure over a space of possible physical laws of the universe. Our definition involves a sigma algebra
$\Gamma$ defined over a state space whose elements we will call ``space-time events''. For maximal
generality, we don't define that state space any further.
We then define ``laws'' to be special kinds of partitions over the sigma algebra $\Gamma$.
This is our first contribution -- a fully rigorous formalization of how to define probability measures over
laws of the universe.\footnote{Arguably, such a formalization of probability measures over
a space of laws should be central to much of modern cosmology, as well as work in the foundations of physics, Bayesian philosophy of science,
and arguably even to some of epistemology. Nonetheless, it seems that no such fully rigorous formalization has been previously introduced in the literature.}

However, the details of this contribution of ours, and in particular of how we identify elements of such partitions 
of $\Gamma$ as possible laws of the universe, do not
directly arise in our analysis of the Boltzmann brain hypothesis. Accordingly, 
the details of our definition are consigned to Appendix \ref{sec:app_A}. The
reader who wishes to confirm that we do not run afoul of these restrictions on
the sigma algebra of $\Gamma$ are invited to consult that appendix.


With this issue resolved, we have the machinery to 
reason from current data to conclusions concerning the 
laws of the universe in a , i.e., to investigate
posterior distributions over laws given various data
sets, e.g., ``$P(\mbox{Newton's laws hold} \,|\, \mbox{observational data } D)$''. In this paper we will mostly be interested in such
observational data sets $D$ that comprise the current state of multiple ``scientific records'' stored in some of our books, 
academic papers, computer memories, etc.
Central to our analysis will be the suppositions (properly formalized via information theory) that there is high probability that every such record
that we consider accurately provides the results of experiments that take place at some time other than the present.

We label the first such observational data set we are interested in as ${\cal{D}}_1$. ${\cal{D}}_1$ consists of current scientific records 
which
lead to the conclusion that our universe obeys Newtonian mechanics
(assuming those records do in fact provide the result of the associated experiments, which are in our past). In other words, they are the current
state of some physical  ``memory system'' that we suppose (!) is statistically correlated with the state of some experimental
outcomes in the past~\cite{wolpert2024stochastic}, where those experiments involved Newton's laws.
Stated more formally, if $\D_1$ --- the \textit{current} state of an associated physical memory system ---
does in fact accurately provide the results of experiments conducted in
\textit{our past}, then $P(\mbox{Newton's laws hold} \,|\, \mbox{observational data } \D_1)$ would be high.
(In this paper we will not need to precisely quantify what it means for some probability to be ``high''.)
Below we will refer to this situation as $\D_1$ being ``reliable''.
%
%

Unfortunately though, just by itself, $\D_1$ cannot establish that
it ``does in fact accurately provide the results of experiments conducted in
our past with high probability''. As a prosaic example of this problem, no set of lab notebooks with squiggles in them can, by themselves, imply that those 
squiggles record the results of experiments in the past concerning Newtonian
mechanics that were actually done---ultimately, one must simply assume that they
have such implications with high probability.
No \textit{a priori} reasoning can allow us to conclude that any such $\D_1$ causes $P(\mbox{Newton's laws hold} \,|\, \mbox{observational data } \D_1)$ to be high. 

To get past this roadblock we need to in essence \textit{assume} that $\D_1$ does indeed
accurately provide the results of past experiments.
More precisely, we need to adopt a prior distribution
over $\Gamma$ that 
says that we can indeed use the records ${\cal{D}}_1$, which are the current state of an associated 
memory system, to infer that with high probability the results of those earlier experiments, which in turn imply Newton’s laws.
While these details don't concern us here, it is important to note that can make this
reasoning concerning prior distributions and our data fully formal. To do so, in 
the terminology of~\cite{wolpert2024memory} we say that ${\cal{D}}_1$ is \textbf{reliable} if it is the current state of a memory system, and there is high restricted mutual information between ${\cal{D}}_1$ and the results of those particular experiments in our past which are related to Newton's 
laws.\footnote{Formally,
``reliability'' is an assumption about the nature of the physical process
coupling the state of the memory system that contains the value $\D_1$ and the state of (some variables in) the universe
in the past. $\D_1$ is, strictly speaking, a value
of the random variable given by the current state of the memory system,
and reliability
is a property of the coupling. 
}

It is important to emphasize that \textit{every} paper concerning the second law, the past hypothesis, or
Boltzmann brains, has assumed the laws of physics hold. Many of those papers go on to make
additional assumptions as well. Here we are maximally conservative, not making any extra assumptions at all.

As shorthand, below we will say ``our prior is that the data ${\cal{D}}_1$ is reliable'' if the combination of our prior over
the set of space-time events in $\Gamma$, together with the likelihood function over $\D_1$
conditioned on $\Gamma$, gives a high posterior probability that ${\cal{D}}_1$ is in fact reliable.
Under such circumstances, we have high posterior probability that the contents of $\D_1$ accurately
reflects the results of earlier experiments. From now on, unless explicitly stated otherwise, we will assume that we always have such a data set $\D_1$.

\subsection{The entropy conjecture}

Now that we have elaborated how  a data set providing the current state 
of a memory system can lead us to believe Newton's laws,
we can start to investigate the thermodynamic implications of those laws.
One of the myriad contributions of Boltzmann was, of course, his argument that the entropy conjecture 
follows from Newton's laws. (More precisely of course, he argued that Newton's laws led to the H theorem;
the entropy conjecture can be viewed loosely as a generalization of that theorem.)
In modern language, his argument concerns stochastic processes, Markov processes in particular.
(See~\cite{lawler2018introduction,shalizi200736} for background on stochastic processes.)
However, Boltzmann derived his results well before Kolmogorov formalized the notion of stochastic
processes, and in particular before anyone had started to systematically investigate Markov processes. As a result,
even though his mathematics behind his analysis was correct, his
formulation of those results does not explicitly involve stochastic processes. 

Unfortunately, this precise failure to formulate arguments concerning the dynamics of 
the universe's entropy as a stochastic process has led to widespread failure to fully grasp the mathematics of the 
entropy conjecture,
and how to use it to reason formally about the second law, BBs, etc.
To undertake such reasoning one must formulate the entropy conjecture in terms of 
stochastic processes.

To that end, we now present a generalized version of the 
entropy conjecture, as a specific class of Markov processes. 
While we will use the term ``entropy'' to mean the random variable of this process, the
underlying mathematics of this kind of process is more general.
Write this Markov process as $\XX$, defined over a 
(compact, real-valued) state space $X$. We write $x_{t}$ for the value of that process at the time $t$, i.e., the entropy of the universe then.
(Note that while $X_t$ the state space of a random variable, $t$ is an index, distinguishing instances of that random variable.)
Refer to a distribution conditioned on $n$ time-indexed entropy value as an ``$n$-time-conditioned'' distribution.
For example, $P(x_3 \,|\, x_1, x_2)$ is a 2-time-conditioned distribution over $x_3$.

We will say a process is a \textbf{Boltzmann process}
if it has a time-symmetric and time-translation invariant one-time-conditioned distribution:

\begin{equation}\label{DistributionSymmetries}
\begin{array}{cc}
& 
  \begin{array}{cc}
   P(x_{t+k}= a\,|\,x_{t}=b) \;=\; P(x_{t-k}= a\,|\,x_{t}=b)\\
   P(x_{t+k}= a\,|\,x_{t}=b) \;=\;P(x_{t}= a\,|\,x_{t-k}=b) \\
  \end{array}
\end{array}
\end{equation}

\noindent where $a,b \in X$ and $k>0$ is arbitrary. The first equality expresses time symmetry (also called time reversal invariance) about a particular time $t$. The second expresses time translation invariance. The particular stochastic process derived in
Boltzmann's H theorem can be viewed as a kind
of Boltzmann process. With $k$ implicit, we refer to the conditional distribution in~\cref{DistributionSymmetries} 
as the \textbf{kernel} that generates the Boltzmann process.  (Because of the time-translation and time-symmetric invariances of the process,
the arguments of the kernel are not indexed by any specific values of time.)

The definition of a Boltzmann process involves probability distributions conditioned on
values of the process at single times only.
However, in many situations we would want to compute the conditional probability of the value a Boltzmann
process $x$ conditioned on its value at two times. In particular, in this paper we will be interested in $P(x_{t}\,|\,x_{t_0},x_{t_f})$
for times $t$ such that $t_0 < t < t_f$.
Fortunately, as is shown in~\cite{scharnhorst2024boltzmann}, for Boltzmann processes
we can express that
two-time-conditioned distribution purely in terms of the kernel of the process:

\begin{equation}\label{Lemma1ProofPart2}
\begin{split}
&\!\!\!\!P(x_{t}=c\,|\,x_{t_0}=a,x_{t_f}=b) \\
&\qquad\qquad=\frac{P(x_{t_0}=a\,|\,x_{t}=c) \,P(x_{t}=c\,|\,x_{t_f}=b)}{P(x_{t_0}=a\,|\,x_{t_f}=b)}
\end{split}
\end{equation}
\noindent for all $t \in [t_0, t_f]$.

We refer to this result as the ``Boltzmann generation lemma''. Note that due to Markovanity, conditioning on an additional value $x_{t_i}$ has 
no effect on this conditional distribution if $t_i$ lies outside the interval $[t_0,t_f]$. So the lemma provides
a broad suite of $n$-time conditioned Boltzmann distributions, for $n$ arbitrarily large.\footnote{Note that the relationship between Boltzmann processes and the kernel 
that generates them is essentially identical to the relationship between (one-dimensional)
Gaussian processes and their (Gaussian) distribution kernels.
In particular, like all other stochastic
processes, both Boltzmann and Gaussian processes are probability distributions defined in terms of a countably
infinite set of random variables (whose index is called ``time'' in the case of Boltzmann processes). This means that 
for both of them we can evaluate distributions conditioned on arbitrary finite sets of indexed values from their state space.}

%
As an aside, an important property of Boltzmann processes is
that their marginal distribution must be the same at all times $t$
i.e., $P(x_{t})$ must be independent of $t$. (This follow from the
fact that the kernel is time-translation invariant.) It immediately follows that
that marginal distribution equals the fixed point distribution of the kernel~\cite{scharnhorst2024boltzmann}.

As implied in the discussion above, the standard argument for the BB hypothesis involves Boltzmann processes
that are conditioned on $S_{t_0}$, a single, specific value of entropy at a single, specific moment in time (namely, $t_0$ is the present). 
As we describe in detail below, successive stages
in that standard argument all involve conditioning on that pair $(S_{t_0}, t_0)$. 
In particular, even if we exploit Bayes' theorem to condition on other quantities, those other quantities
we condition on are all \textit{in addition} to conditioning on $S_{t_0}$.

Accordingly, in our analysis of the BB hypothesis we will focus on (Boltzmann)
stochastic processes that are ``pre-conditioned'', to always be conditioned on some specific set of events $A_1$,
whatever other set of events $A_2$ they might be conditioned on as well. It will be convenient to introduce
some special terminology to refer to such stochastic processes. 

Let $\XX$ be a stochastic process over a state space $X$, and  let $\E := \{ x_{t_1}, x_{t_2}, \ldots \}$ be an associated
set of events.
We will say that a different stochastic process $\XX^{\E}$ is \textbf{equivalent} to the Boltzmann process $\XX$, \textbf{nailed}
to the \textbf{(nailed) set of events} $\E$, if
\eq{
P_{\XX^{\E}}(x_{t'_1}, x_{t'_2}, \ldots ) = P_{\XX}(x_{t'_1}, x_{t'_2}, \ldots \, | \, \E) 
\label{eq:equivalence_def}
}
for all sets of times $\TT' = \{t'_1, t'_2, \ldots\}$ and associated set of joint states $(x_{t'_1}, x_{t'_2}, \ldots, )$.
In other words, 
the unconditioned distribution over joint states $(x_{t'_1}, x_{t'_2}, \ldots, )$ under the process $\XX^{\E}$
is the same as that of the Boltzmann process $\XX$ at those times conditioned on
$\E$. 
Intuitively, we ``nail down'' some of the time-state pairs of the underlying Boltzmann process $\XX$
to produce $\XX^{\E}$. 
It immediately follows from the definition~\cref{eq:equivalence_def} that for all 
sets of events $\E'' := \{x_{t''_1}, x_{t''_2}, \ldots \}$,
\eq{
P_{\XX^{\E}}(x_{t'_1}, x_{t'_2}, \ldots \, | \, \E'') = P_{\XX}(x_{t'_1}, x_{t'_2}, \ldots \, | \, \E, \E'') 
}


%
%

In the sequel, especially when discussed nailed data sets,
it will sometimes be convenient to rewrite the pair of an index $t$ and associated value $x$ as a pair
$(t, x)$, or even abuse notation and writing it as $(t, x_t)$.
As shorthand, we will say that a stochastic process over a state space $X$ is \textbf{singleton-equivalent} if it is equivalent to
a Boltzmann process nailed to a single event, $x_t$. 

As an important example, the so-called ``past hypothesis (PH)'' is the informal idea that
the second law of thermodynamics can be derived from the entropy conjecture if we condition the Boltzmann
process of the universe's entropy on the value of the universe's entropy at the time of the Big Bang~\cite{reichenbach1991direction,wolpert1992memory,davies1977physics,rovelli2022memory}.
Stating this more formally, the PH involves a singleton-equivalent stochastic process, where the singleton concerns
the time of the Big Bang. (This description of the PH is elaborated in a fully formal manner below.)



We now have the tools to precisely state what Boltzmann did show --- and what he did not.
Stated formally, what Boltzmann showed was that if Newtonian mechanics holds, then
the H function evolves as a Markov process having
a time-symmetric and time-translation invariant kernel, i.e., as a Boltzmann process.
So the assumption that ${\cal{D}}_1$ is reliable tells us that
the stochastic process governing the dynamics of H is a Boltzmann process. 
{This clarification of how to properly formalize the entropy conjecture is our second contribution.} 

Note that Boltzmann did \textit{not} prove that
the H function evolves as some specific singleton-equivalent stochastic process. In particular, he did not prove that
it evolves as a Boltzmann process nailed to a singleton data set $S_{t_0}$ where $t_0$ is the present. Nor did he
prove the PH, that the H function evolves as a Boltzmann process nailed to a singleton data set $S(t_{PH})$ where $t_{PH}$ is the Big Bang or some such.
{This further clarification of what the entropy conjecture does \textit{not} say is our third contribution.} 

We need more than just a fully formal entropy conjecture to analyze the BB hypothesis though. 
To properly understand that hypothesis
we also need a properly motivated basis for determining just what nailed data set we should use.
To do this, we can invoke the copious literature on desiderata motivating
Bayesian reasoning. Many compelling arguments, ranging from DeFinetti's Dutch book
arguments~\cite{pettigrew2020dutch} to Cox's axiomatization of a ``calculus of reasoning under uncertainty''\cite{van2003constructing}
to the derivation of subjective Bayesian decision theory based on Savage's axioms~\cite{karni2005savages}, all
establish the normative rule that whenever we perform scientific reasoning, we should only consider probability distributions that are conditioned on precisely all 
of the current observational data that we have, and not on anything more that that. 
%
%
(See~\cite{fishburn1981subjective} for a comprehensive though now somewhat dated review of the relevant literature
on Bayesian reasoning.)

This normative rule is a formalization of what we meant above when we said ``all evidence can be doubted, but doubting selectively leads to shaky conclusions." In the context of the introductory argument, whether or not we conditioned on reliability determined whether or not we could negate the BB hypothesis. Here, we emphasize that, if assuming a prior of reliability, \textit{all} data assumed reliable needs to be conditioned on. In other words, the normative rule implies that we should include in the nailed set all
time-entropy pairs that we have high belief in, and no others.
We will refer to this injunction as the \textbf{(nailed set) construction rule}.

\subsection{The Boltzmann Brain hypothesis and the entropy conjecture}\label{BBHTheorem}

We now have the machinery to rigorously investigate the BB hypothesis. To begin, we 
define the \textbf{BB hypothesis} formally as a triple of suppositions:

$ $

\begin{enumerate}
\item[i)] The entropy conjecture holds for a kernel generating a Boltzmann process $\XX$;

\item[ii)] The stochastic process $\SSS$ giving the entropy values of the universe is equivalent to $\XX$ nailed to 
the singleton $S_{t_0}$,
where $t_0$ is the present and $S_{t_0}$ is the universe's entropy at present.

\item[iii)] The expectation value of the stationary distribution of $\SSS$ is far greater\footnote{The informal 
term ``far greater" is used extensively in the literature to describe the BB hypothesis, the second
law, and the PH. A possible formalization of
 ``far greater" is $P\left(\dfrac{d\SSS}{dt}\Big|_{t_0}>0\right)(S_0)\sim1$. } than $S_{t_0}$.
\end{enumerate}

$ $

\noindent The immediate implication of the BB hypothesis would be that entropy increases going into
our past, i.e., that we are BBs.

As pointed out above,
Boltzmann proved that the first of the suppositions defining the BB hypothesis
follows from Newton's laws (and therefore from $\D_1$). However, the other two suppositions are just as crucial to
have a ``BB'' as the term is usually considered. 
For current purposes it is safe to simply  \textit{assume} that item (iii) of the BB hypothesis holds, by incorporating that
assumption into our prior. (While cosmological arguments indicate that that assumption holds,
there are subtleties that are irrelevant for current purposes.)

Item (ii) of the BB hypothesis is far more problematic however. The entropy conjecture does 
\textit{not} specify that the stochastic process generating the trajectory of entropy values of our universe is singleton-equivalent. 
So the posterior probability of the laws of the universe conditioned only on the data ${\cal{D}}_1$ is \textit{not} peaked about the BB hypothesis,
as that hypothesis is defined above.
%
This clarification of what the entropy conjecture does (not) imply concerning the BB hypothesis is 
our fourth contribution.

In light of this clarification, is there any way we might 
infer the BB hypothesis from the assumption only that ($\D_1$ is true and therefore) the entropy conjecture holds,
as the standard arguments for the BB hypothesis in the literature claim to do?
To answer this question, first define ${\cal{D}}_2$ to be some set of scientific records 
which have the property that {if} the entropy conjecture holds (i.e., if the first supposition of the BB hypothesis holds), and ${\cal{D}}_2$ 
is reliable, then the other two suppositions of the BB hypothesis hold with high posterior probability, conditioned on both $\D_1$ and $\D_2$.
%
We will use the term ``standard argument'' for the BB hypothesis
to mean the assumption that there is such a $\D_2$, and so suppositions (ii) and (iii) of the BB hypothesis hold.

Unfortunately, whether or not we have such a $\D_2$,
we definitely do have yet another, third data set, $\D_3$, which (via the nailed set construction rule) contradicts
supposition (ii) of the BB hypothesis.
It's convenient to partition this set of data $\D_3$ into two distinct subsets, $\D_3(1)$ and $\D_3(2)$,
and discuss them separately. 

First, whatever else we might infer concerning 
the entropy of the universe at various times, we certainly have at least as much data leading us to infer the value of the universe's
entropy at the present, $S_{t_0}$. $\D_3(1)$ is that data leading us to conclude
that the entropy at the present is that particular value. (More formally, we
adopt a prior such that $\D_3(1)$ is reliable, where $\D_3(1)$ implies that equality with high probability.) 
%
Invoking the construction rule, this means that the singleton $S_{t_0}$ should be added to the nailed set.

The second component of $\D_3$, $\D_3(2)$, is a set of data 
which have the property that {if} the entropy conjecture holds (and assuming a prior such that $\D_3(2)$ is reliable),
that set of data would imply that at the time $t^*$ of the Big Bang, or shortly thereafter, the
universe had 
some specific entropy value $S^*$ that is far lower than $S_{t_0}$. A very large number of such data sets
are provided by modern observational cosmology. Invoking the construction rule, this means that the pair $(t^*, S^*)$
should also be added to the nailed set, in addition to $(t_0, S_0)$. This implies that our nailed data set should have
two time-indexed entropy values in it.

As we describe below, 
depending on the precise parameters of the associated Boltzmann process and the
precise values in this two-moment nailed data set, the associated posterior expectation of the values of the universe's entropy 
would not monotonically increase as we move further into our past from the present.
So at least for those precise values in $\D_3$, the three suppositions
that jointly lead to the BB hypothesis would not all hold. Indeed, $\D_3$ would not just 
contradict the usual argument establishing the BB hypothesis; it would in fact rule out the
BB hypothesis. (See Fig.\,1.) 
\begin{figure}[h]
\label{fig:two-time-moment}
\hspace{-0cm}\includegraphics[scale=.6]{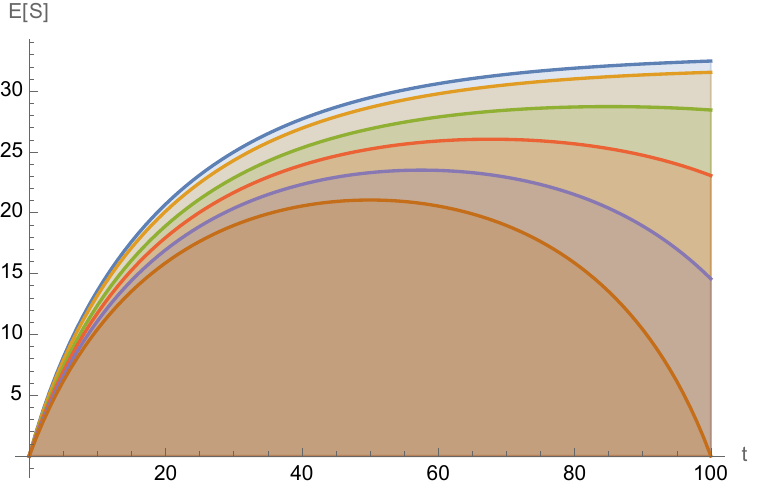}
\caption{An example of a two time conditioned Boltzmann process, with different values for the final entropy condition. For details on the dynamics, see \cite{scharnhorst2024boltzmann}. Note that there is a (attenuated) second law for final conditions close to the relaxation curve, and a decreasing entropy at the final time for conditions substantially lower than the relaxation curve.}\label{BridgeSample}
\end{figure}

As an aside, suppose that before the advent of observational data indicating the existence of
the Big Bang in our past, we did {not} have the data set $\D_3(2)$, only $\D_3(1)$. Suppose that in those times
before Hubble, scientists had our current understanding of the Big Bang, the PH, why memory is temporally asymmetric, etc., 
Arguably, if that had been the case, then
those scientists \textit{should} have concluded (absent other arguments) that we were Boltzmann
brains. See~\cref{sec:counter-intuition}.

\subsection{The Past Hypothesis and the Second Law}\label{PHandsecondLaw}

Does the kind of analysis given above provide a reason to doubt the PH?
What about the second law itself? 
The answers, respectively, are ``yes'', and ``depends what you mean''. 

To see this, first we introduce a fully formal definition of the PH,
just like our formal definition of the BB hypothesis, as a  set of 3 logically independent suppositions:

$ $


\begin{enumerate}
\item[i)] The entropy conjecture holds for a kernel generating a Boltzmann process $\XX$;
\item[ii)] The stochastic process $\SSS$ giving the entropy values of the universe is equivalent to $\XX$ nailed to 
the singleton $(t^*, S^*)$, where $t^*$ is the time of the Big Bang (or some specific time shortly thereafter) and $S^*$ is the entropy of the universe at the entropy of the universe at $t^*$.

\item[iii)] The expectation value of the stationary distribution of $\SSS$ is far greater than $S^*$.
\end{enumerate}
Note that this formal definition of the PH is almost identical to the formal definition 
of the BB hypothesis, with only the second of the three assumptions being (slightly) different
in the two cases.

$ $
%
%
To begin, in direct analogy to the case with the BB hypothesis, simply {assume}
that item (iii) of the PH holds, by incorporating that
assumption into our prior. 
Next, suppose we had data and adopted associated priors that led us to
conclude that item (i) of the PH holds, and that the nailed data set should include $(t^*, S^*)$. For example, this could
be because we have data $\D_1$ and
$\D_3(2)$, and assume that both are reliable. Suppose though that we did not have any data $\D_3(1)$.


If we made no other assumptions and had no other data, then by the construction rule,
we would conclude that the stochastic process
of the values of entropy was a Boltzmann process singleton-equivalent to $(t^*, S^*)$. In other
words, it would establish item (ii) of the PH, to go with items (i) and (iii). So the PH would hold in its entirety.

That in turn would lead us to conclude that the values of entropy
between the Big Bang and now was monotonically increasing in expectation value, and that
it would continue to increase into our future. (More precisely, that
data together with those assumptions
would mean that the expected value of the marginal distributions of that process at single times $t$ increased monotonically 
with $t \in [t^*, t_0]$.) 

So a prior that $\D_1$ and $\D_3(2)$ are reliable, {together with the assumption that
item (iii) of the PH holds}, \textit{and no other data}, would imply that the PH holds, and that therefore the second law holds.
Note that the second law would in turn
mean that entropy decreases the further back we go into our past from the present. However, as mentioned
above, the BB hypothesis would imply that entropy \textit{increases} going backward in time from the present.
So the PH would both imply the second law and rule out the BB hypothesis.
This is precisely the scenario considered in~\cref{sec:II}, with $\cal L$ set to $(\D_1, \D_3(2))$.

It is important to emphasize though that in addition to $\D_1$ and $\D_3(2)$,
we have yet other data that we typically assume to be reliable.
In particular, it seems hard to dispute that the value $S_{t_0}$ of the universe's entropy at the \textit{current} time $t_0$ is known 
at least as reliably as
$S^*$, the universe's entropy at the time of the big bang. (In this regard, note that modern cosmology only infers $S^*$ indirectly, through an extremely long
line of reasoning involving lots of experimental data of many different kinds.) 

The pair $(t_0, S_{t_0})$ is precisely the data set given by the first component of $\D_3$,
discussed above when analyzing the formal foundations of the BB hypothesis.
According to the construction rule, our analysis is flawed if we condition only one or the
other of the two pairs $(t^*, S^*)$ and $(t_0, S_{t_0})$; that is
the flaw in the BB hypothesis discussed above.
Note though that this flaw in the standard argument for the BB hypothesis is precisely replicated in 
the PH. Both lines of reasoning violate the construction rule (and therefore contradict
all the desiderata underlying Bayesian reasoning), since they selectively leave out some reliable data 
from the nailed data set while keeping some other data. In fact, arguably the PH is more
guilty of this transgression than the standard argument for the BB hypothesis, since the reliability of the first component of 
$\D_3$ seems more indisputable than that of the second component of $\D_3$. This means that any physicist willing
to accept the PH should instead adopt the BB hypothesis (absent other arguments). 

Of course, it is possible to {combine} the suppositions of both hypotheses,
simply by supposing the full data set $\D_3$, including both subsets, is reliable.
Doing this would lead us to conclude that the evolution of the universe's entropy
is equivalent to a Boltzmann process nailed to the pair $\{(t^*, S^*), (t_0, S_0)\}$,
as illustrated in Fig.\,1.
%
Reference \cite{scharnhorst2024boltzmann} analyzes such Boltzmann processes that are nailed to such a pair
of time-entropy pairs. It is shown there that for certain parameters of the
Boltzmann process and certain values in the nailed data set, the associated posterior
expected
entropy would have been \textit{increasing} as we start to go backwards in time, into our very recent past. 
This would imply that the second law is wrong in our recent past. Expected
entropy would still decrease though once we go far enough backward in time towards the Big Bang. So
the second law would be correct on long enough timescales into our past. Moreover,
we would still have the usual second law going \textit{forward} from the present. So
all experimental tests of the second law we might do \textit{in our future} would (according
to the entropy conjecture, etc.) still confirm the second law.
%
Such a scenario could be viewed as a highly attenuated variant of the BB hypothesis, 
one not necessarily having any implications about carefully arranged 
sensory inputs to our brains and / or carefully arranged photons streaming in to our telescopes, etc. 
%
%

\subsection{What the entropy conjecture does not imply}\label{DoesntImply}

Interestingly, the second law provides a major way -- perhaps the \textit{only} way, in fact -- for us to 
conclude that any scientific records are reliable. This is because of the role of the
second law in establishing reliability~\cite{rovelli2022memory,reichenbach1991direction}, or stated more formally, 
because of the role of the second law in real-world type-3 memory systems~\cite{wolpert2024memory}.
So assuming the second law holds provides the only way we know of to conclude that $\D_1$ is reliable, along with 
establishing the reliability of either subset in $\D_3$, or even some $\D_2$. Given the results of~\cref{PHandsecondLaw}, this means
that  the only known way to establish that the second law holds is, ultimately, to assume that the second law holds.

So the second law is consistent with the priors that imply it --- but only by virtue of the fact that
we can adopt priors that ${\cal{D}}_1$ and $\D_3$ are reliable,
or that the second law holds (or even both). But we cannot somehow avoid making any prior assumptions
at all, in an attempt to bootstrap ourselves into placing high probability on having ${\cal{D}}_1$ or $\D_3$ be reliable and
that the second law of thermodynamics holds.  (See~\cref{app:circular} for a Bayesian formalization
of circular reasoning, and associated derivation of what circular reasoning does and does not imply.) 

Suppose that we misunderstood the foregoing, and thought that somehow if $\D_1$ were reliable, that would
not only imply the entropy conjecture, but also imply: 
\begin{enumerate}
\item [a)]
\label{item:1} The associated evolution of the entropy of the universe is
a Boltzmann process that is singleton-equivalent, for a time $t_0$ set to the present, with $S_{t_0}$ set
to the entropy of the universe of the present;
\item [b)]
\label{item:2} If the second law holds
(both at the present and at times in our past), then it follows that in fact $\D_1$ is reliable. (In other words,
there is a type-3 memory system that would rely on the second law in order to establish that $\D_1$ is reliable.)
\end{enumerate}
Suppose as well that the second law is true.
That supposition would mean (by (b)) that $\D_1$ is reliable, and therefore Newton's laws hold. This in turn would mean
(as Boltzmann showed) that the entropy conjecture holds. By (a), we could then conclude that the BB hypothesis holds.

However, this derivation of the BB hypothesis would mean that the second law does not hold. That would contradict the
second supposition (b) of this very derivation of the BB hypothesis. Note that this is more than using
circular reasoning to try to establish some hypothesis or other. Rather it is using reasoning that contradicts itself
to try to establish a particular hypothesis. 
%
This seeming self-contradiction is a version of unstable reasoning; see~\cref{sec:app_B}.

Of course, the foregoing flaw in a particular derivation of the BB hypothesis
does not mean that the BB hypothesis is false --- it just means that the reasoning 
for that hypothesis given above contradicts itself.  More precisely, as we prove
in~\cref{app:circular}, unstable reasoning just means that one is implicitly trying to use a joint prior
distribution over all associated random variables that is not a proper probability distribution.
So it's simply saying that one needs to adopt a different prior. This clarification that issues
of unstable reasoning have no bearing on the BB hypothesis is our fifth contribution.



%
%
%


In addition though, it is interesting to note that \textit{nobody} has ever invoked the
second law as in (b) to justify the reliability of any data concerning the past they wish
to use, be that data ${\cal D}_1$ or anything else. Rather, we now understand that
all memory systems that allow us to treat the current state of a memory
as providing information concerning the past must,
ultimately, rely on the second law.\footnote{Indeed, when the memory system in question
is the human brain, we have no idea of the details of how the second
law establishes the validity of memory. We just believe that this must be the case.}
But before this understanding, say a century ago, there would have been no reason
to suppose that the second law was in any way required for us to (have faith that we can) use
our experimental data as though it provides accurate information about the past state
of our experimental apparatuses. There would not have been any reason
to suppose that the second law is itself in any way needed to derive the second law itself. 
And no current textbook uses the second law that way. ``Unstable
reasoning'' never explicitly arises in any of the arguments for the BB hypothesis
one can find in the literature.

Finally, it is important to emphasize that there is a simple variant of the
BB hypothesis that does not suffer from any form of ``cognitive instability''.
Note that formally speaking, the analysis in~\cref{sec:II}  says that the BB hypothesis defined
as a Boltzmann process singleton-equivalent to the present time, when $\cal D$ is observed, does not hold
(if we assume $\cal R$). Moreover, it actually also implies that the dynamics of the universe's
entropy is equivalent to a Boltzmann process nailed to a pair of times, namely the times
of $\cal D$ and $\cal P$. However, as discussed above, this could in turn imply that \textit{entropy
increases going into the past from the time of $\cal P$}, when we started our experiments that
led us to believe $\cal L$. In
other words, that argument not only says that the present is not a Boltzmann brain -- it also
says that the time when we started our experiments \textit{is} a Boltzmann brain, and that
entropy increases going into the past of that time, as well as going forward from it, to our present.

It is illuminating to push this scenario further, by introducing a slight variant of the BB hypothesis.
This variant is just like the standard BB hypothesis in that it supposes that the universe fluctuates from a state
of maximal entropy down to some minimal entropy value, after which the entropy of the universe starts expanding again.
The difference is that rather than suppose that the minimum of the entropy of the universe occurs at the present, under the
1000CE BB hypothesis it occurs at 1000CE.\footnote{Such a scenario has sometimes been considered informally in the literature, where
it has been called a ``Boltzmann bubble" \cite{ChenSelfLocating, WallaceUndermining}. }
More formally, the ``1000CE BB hypothesis'' is
the same three suppositions that were used to define the BB hypothesis in Section \ref{BBHTheorem},
except that $t_0$ is changed to be the year 1000CE:

$ $

\begin{enumerate}
\item[i)] The entropy conjecture holds for a kernel generating a Boltzmann process $\XX$;

\item[ii)] The stochastic process $\SSS$ giving the entropy values of the universe is equivalent to $\XX$ nailed to 
the singleton $(t_{1k}, S_{1k})$,
where $t_{1k}$ is 1000 CE and $S_{1k}$ is the universe's entropy at that time.

\item[iii)] The expectation value of the stationary distribution of $\SSS$ is far greater than $S_{1k}$.
\end{enumerate}

$ $

Under the 1000CE BB hypothesis, the second law would hold, so our memory systems (scientific records) would
accurately record the results of all experiments conducted in the last thousand years. 
So the data set $\D_1$ which implies the entropy conjecture would be reliable. So cognitive instability arguments don't apply, 
even though a version of the BB hypothesis is still true.\footnote{There 
are other problems with the criticism that the justification of the BB hypothesis based on (a), (b) is unstable.
Most obviously, these criticisms make their case by first
seemingly proving some particular proposition $A$ based on some particular data (in out case, proving the entropy conjecture from $\D_1$ via Newton’s laws).
They then seemingly show that 
$A \rightarrow B$ (in our case $B$ is the BB hypothesis). They end by seemingly showing that $B$ contradicts $A$.
From that seeming contradiction they conclude that $B$ must be wrong.
That conclusion is fallacious; unless one can show explicitly that $A \rightarrow B$ is not
logically sound, then the contradiction means that the premise 
$A$ must be wrong, with no implications for whether $B$ is wrong. In our case, this would mean 
that the contradiction proves that Newton's laws are wrong. (!)}  
%



Note that we could keep pushing the time of the entropy minimum
further and further back into the past, from the present to 1000CE and ultimately all the back to the big bang. At that point 
we have a variant of the BB hypothesis which is just the fluctuation hypothesis version of the PH \cite{ChenSelfLocating}. 
Moreover, as discussed above, we \textit{do} have a data set -- $\D_3$ -- that we think gives
us some confirmation that entropy was indeed quite low at the big bang. So one must be careful before making sweeping
claims about whether any particular variations of the BB hypothesis could or could not be confirmed by data,
since the PH is itself only such a variant.
These formal definitions of the BB hypothesis, the PH, and the 1000CE hypothesis are our last contribution.

As a final comment, as pointed out above, this same kind of data problem 
holds for the second law, the PH, etc., in that none of them can
be either confirmed or refuted \textit{solely} from data. Any data-based assessment of any of them ultimately relies
on making assumptions for prior probabilities. In our analysis in this paper, we have chosen to only adopt the
prior that $\D_1$ is reliable, which we take to be a necessary minimal assumption for \textit{any}
investigation of these issues.\footnote{In particular, we have been careful \textit{not} to introduce any
assumptions about what the nailed set should be when we use the entropy conjecture. The second law,
the validity of scientific records, the PH, the BB hypothesis (in all of its invariants) all make some
assumption for that nailed data set, without any formal justification in terms of Bayesian decision theory.} 
Of course, this prior removes the need to invoke
the second law to establish the reliability of $\D_1$. So there is no cognitive instability under this prior.

\section{How and why intuition leads us astray}
\label{sec:counter-intuition}

Many of the mathematical arguments and derivations presented above defy 
to common sense and intuition. In part, one might dismiss this issue. After all,
since the advent of quantum physics and the general theory of relativity,
physicists have become accustomed to following math wherever it may
lead, even when it results in counterintuitive conclusions.

Nonetheless, it is worth teasing apart the precise reason \textit{why} our
result are counterintuitive, to allay associated fears.
Some instances of specific intuitions being defied were discussed above,
in both the introduction and~\cref{sec:III}. Since these are
such subtle points however, It's worth describing some of the other strong intuitions
which are at odds with our formal results, why those intuitions
are so strong, and just how it is that they can be so misleading.

First, most obviously, almost everyone \textit{feels}, very deeply, that
their memories are accurate, that they correctly depict
some information about a real external world and its states in the
recent past. To understand how this deep-felt intuition could
be completely wrong, it's necessary to proceed through several
steps, to understand the source of this intuition:
\begin{enumerate}
\item To begin, note that human memory is, formally speaking, a
physical system (namely, our brains) whose current state provides statistical information
about the likely state of some external physical system, at  a different time (namely, information
about the external world that we believe we remember). Or at least, we presume
that to be the case.
 
Formalizing memory this way, we immediately notice a mystery: why do we have so much
confidence that the current state of our brains provides such
accurate information about the \textit{past} state of the external world,
but not about the \textit{future} state of the external world? Phrased differently, why
do we believe our innate ability at retrodiction is so much greater than our ability
at prediction? 

Presumably, our belief in the accuracy of our memories of the past does not have anything to
do with our knowledge of the laws of physics. Training in theoretical thermodynamics
is certainly not required for people to believe their memories are accurate (!).
Rather, this belief seems to be built into the functioning of our brains, into how our
neocortices are wired, so to speak.
Why are our brains constructed this way, to believe our memories of our past, solely?

\item To address that mystery requires us to provide a physical model 
of memory. However, human memory involves some extraordinarily complicated
processes in the human brain, processes we know precious little about.\footnote{We 
know for example that the hippocampus plays a crucial role, and that
elaborations of Hebbian learning, like post-synaptic plasticity, play a major
role at the level of individual neurons. But even this much is understood only
in generalities.}

Accordingly, the great majority of investigations of the human memory system have not
considered it directly. Rather they have considered far simpler examples of memory systems that 
we can investigate in detail, which provide non-biological instances of
memory~\cite{wolpert1992memory,wolpert2024memory,rovelli2014we,rovelli2022memory,reichenbach1991direction,davies1977physics}. 
A common example of such a memory system is a the surface of the moon.
The surface of the moon is the memory system, akin to the human brain,
whose current state (namely, the craters on the moon's surface) provides information about
the state of physical systems external to it in the past (namely, that there were meteors 
on a collision course with the moon in the past). Another common example is
a line of footprints on an otherwise smooth beach. The beach 
is the memory system, akin to the human brain,
whose current state (namely, the footprints on it) provides information about
the state of physical systems external to it in the past (namely, that there was a person
walking across that beach). 

\item What do we learn from these examples of simple memory systems concerning the asymmetry of our memory? Do 
the common features of these
examples of memory systems help us understand why we have so much more confidence
in retrodiction rather than prediction? 

In fact, these examples do have one
striking feature in common: all of them seem to rely on the second law of
thermodynamics. The temporal asymmetry of these memory systems, and by extension
of our human brains, seems to reflect nothing more than the simple fact that entropy
increases in time. Indeed, the precise reason we are so very sure about some
of our memories seems to reflect the ironclad nature of the second law.

In other words, the reason we seem to have so much confidence in our memory 
ultimately resides in the fact that those memories rely on the second law.

\item This line of reasoning would seem to mean that we \textit{should} have complete
faith in our memories, given that the second law cannot be violated. Unfortunately
though, this line of reasoning begs a question: how it is that the
physical universe, whose (relevant) microscopic laws are all time-reversal symmetric,
could give rise to the asymmetry of the second law? The arguments supporting the second law
offered by Boltzmann and those who followed him were all based on the microscopic laws
of physics, and therefore they \textit{had to be} time-inversion symmetric. Formally
they all state that f at a certain time $t$ entropy is low (or a similar quantity is, like Boltzmann's H function),
then entropy increases in time going \textit{both} forward from $t$ and backward from $t$~\cite{davies1977physics}.
This implies that entropy should increase going both forward from the present \textit{as well as going
backward from the present}.

In other words, our analysis of why we believe our memories to be accurate has  reduced
the mystery of human memory to  the infamous paradox ascribed to Loschmidt.

\item What would this time-symmetric reasoning of Loschmidt (and others) mean physically?
What would it mean, physically, for entropy to increase going into our past, despite our apparent
memories of the past, which would seem to rely on entropy increasing in time,
not decreasing? Among other things, it would have to mean that we have no reason to believe
our memories are accurate. 
It means that they do \textit{not} tell us anything about the state of the external world in the past.
They only appear to give us that information.

In short, our intuition concerning our memories leads us greatly astray. In
quantum mechanics, or relativity, that which seems ``obviously'' true is deeply problematized, 
either by the results of experiment (quantum mechanics) and / or
pure mathematical reasoning (general relativity). The same is true of thermodynamics,
evidently.

Just like when we analyze event horizons or the quantum Zeno effect,
when when we analyze the second law, memory and Boltzmann brains, we
are almost guaranteed to go astray if we don't rely solely on mathematical
reasoning to guide us, consigning intuition to a purely secondary 
role. 

\end{enumerate}

Reasonable as the arguments just presented might be, in the
abstract, how, concretely, can they hold? How could we have \textit{all} of our human memories concerning the
past be fallacious? How could entropy increase into our past rather than decrease, as required by
the time-symmetric nature of all derivations of the second law that are consistent
with the microscopic laws of physics? How could it be that our memories are wrong?

Such flaws in our memory would require some exquisite fine-tuning, that all the neurons in our brains
happen to be in the state corresponding to particular memories, when in fact nothing
of the sort is true. Amazingly though, standard arguments of statistical physics
tell us that it is almost infinitely more likely for this to be the case, rather than
for entropy to continue to decrease into our past, as demanded by the second law.

This conclusion is precisely the Boltzmann brain hypothesis. As we show above, 
it too has subtleties that can only be addressed through careful mathematical
reasoning.

\section{Discussion}
In this paper, we extended recent advances in stochastic process theory in order 
to disentangle the second law, the reliability of human memory, the BB hypothesis,
and the Past Hypothesis (PH) with full formal rigor.

To do this we first established the following, preliminary results (some of which are similar to 
earlier results in the literature):

\begin{itemize}
\item [i)] The standard argument in favor of the BB hypothesis 
can be seen as contradicting itself, by implicitly assuming the second law;
\item [ii)] Boltzmann's entropy conjecture, properly formalized, says that the H function evolves 
as a particular type of stochastic process, a ``Boltzmann process'',
and then derive some new, useful properties of Boltzmann processes;
\item [iii)] We point out that Boltzmann's reasoning does \textit{not} prove that the H function evolves as a Boltzmann process
conditioned on the value of H at a single time (any more than he proved that it evolves as a Boltzmann process
conditioned on the value of H at two times, or zero times). This disproves the common
interpretation of what Boltzmann proved;
\item [iv)] We introduced formal definitions involving Boltzmann processes
of the BB hypothesis, the Past Hypothesis (PH), and a variant of the BB hypothesis 
that does not suffer from any form of cognitive instability, which we call the 1000CE hypothesis;
\item[v)] We point out that due to (iii), the entropy conjecture has nothing to say one way or another
about the legitimacy of the BB hypothesis, nor of the second law, the PH, or the 1000CE hypothesis.
\end{itemize}

We then used these preliminary results to disentangle the BB hypothesis, the time-asymmetry of 
memory, the entropy conjecture, and the second law, with more formal rigor than has been done before.
Specifically, we demonstrate that neither the reliability of experimental data nor our possessing Type-3 memory {can} be disentangled from the second law. Second, we show that the validity of the entropy conjecture is independent of the BB hypothesis, the PH, and the 1000CE hypothesis, since it is an independent ingredient in those hypotheses. Third, we show that those hypotheses are all variants of one another, 
simply differing in the times they assume for the extreme of an entropy fluctuation --- with the caveat that
it the BB hypothesis has a seeming self-contradiction baked in.

Our use of stochastic process theory -- never before used to analyze these issues --
was necessary to avoid the implicit assumptions in many arguments in the literature, that often result in circular 
reasoning. Stochastic process theory also (correctly) forces
us to specify the prior probabilities one must use to analyze the issues we consider. 

This rigor demonstrates that, ultimately, the only reason we have for assigning more credence to the PH or even to the
second law than we assign to the BB hypothesis is
the credence we assign to our cosmological observations concerning the big bang. This means in turn that before the early
20th century, when we first made those cosmological observations, we should have treated the BB hypothesis as just
as probable as the second law. Moreover, whatever reliability we now assign to those cosmological observations implying
the entropy at the big bang was low, we should arguably assign at least as much reliability to our observations
concerning the \textit{current} entropy of the universe. Once we do that however, then the
Boltzmann process governing the evolution of the universe's entropy should be conditioned on \textit{two}
moments in time (the Big Bang and the present). However, conditioned that way, the Boltzmann process
of the universe would rule out not just the BB hypothesis, but also the 
naive versions of the PH and the second law currently in vogue.


We emphasize that since the entropy conjecture cannot make a conclusion regarding the probability of the BB hypothesis that doesn't resort to priors (see Appendix), any conclusion that the BB hypothesis should be discarded entirely must 
involve justification of priors. And as stated previously, it seems very hard to justify a prior that allows one to conclude that the BB hypothesis is 
not true. 
There is not a fully rigorous argument, yet, in the domain of physics to dispel the possibility of the BB hypothesis. The view we have established in this paper is that the BB hypothesis is not established by the standard argument, nor is it refuted by standard criticisms. 

It is important to appreciate that there are many deep issues in
the philosophy of science that our careful investigation of the BB hypothesis
uncovers. One is the huge issue of how --- from the perspectives of either philosophy or
the axiomatizations of Bayesian probability --- one can justify conditioning on \textit{any} moment in addition to the present, and if so, why just one such additional moment. 

Suppose that we do though, e.g., due to associated assumptions concerning reliability. This would still leave the
issue of whether (as implied by the
construction rule) we really should
nail stochastic processes to \textit{all} time-entropy pairs that we have very strong reason to believe. All of which
in fact provides ways to rescue the BB hypothesis, by justifying a prior that we nail the Boltzmann process to only $(t_0, S_0)$.


\acknowledgements{Great thanks to Sean Carroll and Wayne Myrvold for many conversations on this topic, and thanks also to Artemy Kolchinsky. 
DHW also thanks the SFI for support.}


\bibliography{/Users/davidwolpert/Dropbox/BIB/refs.main.1.BIB.DIR,refs.main.1.BIB.DIR-2.bib}

\appendix

\section{A formal definition of scientific laws}
\label{sec:app_A}


Here we sketch a mathematical definition of scientific ``laws''. 
This definition will justify formally our use of probability theory to analyze such laws, i.e.,
it will provide the formal foundation underlying the analysis in the main text.
 
There has been lots of consideration of what is meant by a scientific ``law'' 
in the philosophy of science literature, including some work on how to define probability distributions
over such laws~\cite{godfrey2009theory,godfrey2016other,dasgupta2021current,sprenger2019Bayesian,glymour2003Bayesian}. 
In some senses, our definition can be viewed as a variant of the ``regularity'' view of
what ``laws of Nature'' are~\cite{ArmstrongLawofNature} in that philosophy of science literature. However, our approach
is far more formal and mathematically rigorous than almost all of that literature. Moreover,
the elaborate considerations in the philosophy of science concerning laws of Nature
is in many ways orthogonal to our concerns, and
there is no need for us to carefully situate our mathematical definition within that literature.

To begin, suppose we have some set $\Omega$ whose elements we refer to as \textbf{space-time events} (which we don’t define further).
Define a family $E = \{E(i)\}$ of subsets of $\Omega$ indexed by the argument $i$. We call each member of that family, $E(i)$,
a \textbf{universe}. So $E$ is a family of universes,
each of which is a subset of the set of all possible space-time events.

One might think that it would suffice to have our probability distributions defined over $\Omega$.
However, this would not allow us to compare probabilities of different possible universes conditioned on the same space-time event.
Accordingly, while we don’t need to get into the formal details, here we implicitly 
suppose our probability distributions are all defined over $\Gamma := \Omega \cup E$. So the elements
of our sigma algebra are defined over the union of a set of space-time events and a collection
of universes (all of which in turn comprise a subset of that set of space-time events). This will allow us, for example, to
compare the probability of a universe $E(i)$ conditioned on some space-time event $\omega \in \Omega$
to the probability of a different universe $E(i')$ conditioned on the same space-time event $\omega$.


We semi-formally define a ``law'' $L$ that ``holds'' in a universe $E(i)$ as any collection of subsets of $E(i)$,
$L = \{E(i)_{L,j}\}$, where each such subset $E(i)_{L,j}$ is partitioned into two sets: a set of \textbf{determining} (events) and a 
separate set of \textbf{consequent} (events). We will sometimes refer to such a subset of space-time events 
$E(i)_{L,j}$ as an \textbf{instance} of the law $L$ in the universe $E(i)$. 

Intuitively, each $E(i)_{L,j}$ is a set of space-time events that are related to one another by the associated law $L$.
The determining events are those elements in $E(i)_{L,j}$ that are sufficient, under the law $L$, to force the
consequent events in $E(i)_{L,j}$. Note that the same set of space-time events might arise in two different
instance of the law $L$, $E(i)_{L,j}$ and $E(i)_{L,j'}$, since the partition into determining and causing events 
differs. In particular, this is almost always the case with the time-symmetric laws that we are familiar with.


As an example, in our universe, $E(i)$, we would identify Newton’s laws, $L$, as simply all sets of events in space-time, $E(i)_{L,j}$, such that Newton’s third law uniquely fixes the consequent events in $E(i)_{L,j}$, given the joint state of the other, determining events in $E(i)_{L,j}$.
As an illustration, Newton's laws $L$ would have an instance $E(i)_{L,j}$ (identified by one particular $j$) where in
all of the determining events $\omega \in E(i)_{L,j}$, Galileo is dropping a canon ball from the top of the leaning tower of Pisa,
with some timing device next to him,
and in all of the consequent events $\omega' \in E(i)_{L,j}$ Galileo is at the top of the tower watching as the canon ball hits
the ground, with the same timing device next to him. Note that there are many such determining events $\omega$; they differ
in the associated events happening to the other variables in the universe that do not involve Galileo during
the experiment, e.g., events happening in the Andromeda galaxy
when he is running the experiment. The same is true for the consequent events $\omega' \in E(i)_{L,j}$. We do not
impose a requirement for how the states of those variables concerning events happening in the Andromeda galaxy specified in $\omega$
are related to those in $\omega'$; such relationships would be defined by other laws $L' \ne L$.

As another example, what we here mean by a specific law is formalized as a mathematical relation 
in~\cite{tegmark1998theory,tegmark2008mathematical}, between the ``determining'' and ``consequent'' subsets of all instances of that law.
However, our formalization here is in many ways more general than the approach in those papers, which is focused
on formal systems and model theory.

So for us, a universe $E(i)$ is a set of space-time events, where some subsets of $E(i)$
obey one law (e.g., Newton's laws), some subsets obey a different law (e.g., Maxwell's equations), 
etc. In general of course, due to relativistic considerations, we should identify a law with a set of
subsets of $E(i)$, related to one another by changes in the inertial frame. Similar considerations hold
to capture quantum mechanics in our approach. We do not go into the details of such considerations
here. So the formalism as stated suffices for classical scenarios --- exactly like Boltzmann's H theorem.

As an aside, one can suppose that we restrict attention to universes 
$E(i)$ and associated sets of laws, $L(E(i))$, such that every space-time event in the universe $E(i)$ occurs as part of either a determining or consequent set of space-time events in some instance $E(i)_{L,j}$ of a law that holds in $E(i)$. In other words, we could
exclude space-time events that are neither caused by any other events nor cause any other event, as being meaningless.
No such requirements are necessary for the analysis in this paper.

This technical machinery might seem like overkill. However, without some such machinery we cannot meaningfully say that the same law
$L$ exists in two different universes $E(i)$ and $E(i')$. That in turn is necessary for us to meaningfully discuss things like the
 ``probability'' of our universe obeying some law $L$ rather than some other law $L'$.

\section{Circular reasoning and unstable reasoning}
\label{sec:app_B}

In general, a line of reasoning that bounces back and forth between two propositions $A$ and $B$, without
invoking any other propositions, can neither prove nor disprove $A$ and / or $B$. For example, just because
some such reasoning is circular does not mean that either $A$ or $B$ is not true; it simply means that that
line of reasoning, by itself, does not imply anything about the truth of those two propositions.
Similarly, just because a line of reasoning that bounces back and forth between two propositions $A$ and $B$
is ``unstable'' does not mean that either $A$ or $B$ is not true; it simply means that that
line of reasoning does not imply anything about the truth of those two propositions. 

In this appendix we prove this using Bayesian analysis, and indicate the implications
for criticisms of the BB hypothesis that it is based on unstable reasoning.

\subsection{Circular reasoning in analysis of Boltzmann Brains}
\label{app:circular}

Say we have random variables $A$ and $B$, with
$a \in A$ and $b \in B.$ Suppose that their joint distribution
has conditional distributions such that
\eq{P(b\,| \,a) \simeq 1, P(a\,|\,b) \simeq 1
\label{app:eq.1}
}
\noindent
for two specific values $a, b$.
Then, we can conclude that 
\eq{
& \dfrac{P(a\,|\,b)}{P(b\,|\,a)} = \dfrac{P(a)}{P(b)} \simeq 1
}
i.e., $P(a) \simeq P(b)$.

That is \textit{all} we can conclude. We can’t conclude anything about whether $P(a)$ and/or $P(b)$
is large or small -- so long as $P(a) \simeq P(b)$, all values of $P(a)$ and $P(b)$ are possible.
Trying to instead use \cref{app:eq.1} to conclude $P(a) \simeq 1$ is what is commonly called ``circular reasoning".

\subsection{Unstable reasoning in analysis of Boltzmann Brains}

Consider again the standard argument for the BB hypothesis discussed above, based on data sets $\D_1$ and $\D_2$. Suppose
we had such data sets, and therefore concluded that the BB hypothesis holds. 
One might suppose that the BB hypothesis in turn somehow means that $\D_1$ cannot
be reliable, since (under a common interpretation) the BB hypothesis says that our memories of the past are ``illusory'',
not actually reflecting what was happening then, but only fluctuation in our current neurobiological functions.
So it would seem that having both the 
data sets $\D_1$ and $\D_2$ (which would mean that the BB hypothesis holds) would indicate that we cannot in fact have a data set meeting the definition
of $\D_1$. 

This conundrum in our using $\D_1$ and $\D_2$ to infer the BB hypothesis, which then it turn refutes $\D_1$, is an
example of what has sometimes been called \textbf{unstable}
reasoning~\cite{carroll2020boltzmann}. It is highly nontrivial to rigorously
analyze the precise nature of this phenomenon in the context of the BB hypothesis, where issues like non-equilibrium
dynamics, subsystems that are not fully closed from the outside world, etc., highly complicate the
picture. (Some of these complications are briefly touched on below.)
However, we can get some insight by using Bayes' theorem to analyze a minimal model of unstable reasoning. 


To begin, suppose we have data $a$, and so know that $P(A = a) = 1$.
Then using~\cref{app:eq.1} we can conclude $P(b) \simeq 1$. That’s Bayesian reasoning.
Now, suppose that instead of~\cref{app:eq.1} we have the two conditional distributions
\eq{
\label{eq:app_B_2_1}
&P(B = b\,|\,A =a) \simeq 1\\
&P(A = a\,|\,B = b) \simeq 0
\label{eq:app_B_2_2}
}
where $B$ and $A$ are two random variables. Such a pair of conditional distributions is summarized as ``unstable reasoning''.

When we have such reasoning, we can’t conclude anything about the marginal probabilities of our random variables. 
In particular, we cannot conclude that (with high probability) $A = a$ and / or that $B=b$. Similarly,
we cannot conclude that (with high probability) $A \ne a$, nor that $B \ne b$. This is just like the case of circular reasoning,
described in~\cref{app:circular}.

On the other hand, whereas the two conditional distributions investigated in in~\cref{app:circular}
resulted in having $P(A = a) \simeq P(B = b)$, the analogous conditional distributions considered here give
\eq{
\label{eq:app_B_2_3}
&\dfrac{P(A = a\,|\,B = b)}{P(B = b\,|\,A =a)} = \dfrac{P(A = a)}{P(B =b)} \\
& \simeq 0
\label{eq:app_B_2_4}
}
So $P(A = a) \ll P(B = b)$.

Next, suppose that we also assumed that 
\eq{
P(A = a) \simeq 1
\label{eq:app_B_5}
}
Combining that assumption with our conclusion that $P(A = a) \ll P(B = b)$ would mean that
$P(B = b) \gg 1$, i.e., it would result in a contradiction. So 
there is no joint distribution $P(A, B)$ that obeys~\cref{eq:app_B_2_1,eq:app_B_2_2,eq:app_B_5}.
Our supposition for
at least one of those three distributions must be wrong. 

In the context of the BB hypothesis (as formalized in the main text), to begin we could
note that tautologically, all the data we have only \textit{directly} concerns the present. 
Next, one could take $A$ to be the bit of whether 
$\D_1$ is reliable, with $A = a$ being the event that
$\D_1$ is in fact reliable. (Note that while $A = a$ would imply the second law,
 $A \ne a$ would \textit{not} mean that the second
law is false, only that $\D_1$ is not reliable.) $B$ could instead be the bit of whether the BB hypothesis is true,
with $B = b$ being the event that the BB hypothesis is in fact true. More precisely, it could be the bit of whether 
entropy evolves in a Boltzmann process fixed to a low value in the present, and therefore entropy
was higher in the past. 

So~\cref{eq:app_B_2_2} would reflect the fact that
under the BB hypothesis, \textit{everything else being equal} all our (current) memories of the past are likely to be ``fictional'' fluctuations
in the state of of our brains' neurons. This would mean that $\D_1$ is not reliable.

\cref{eq:app_B_2_1}
would instead start from the fact that if $\D_1$ is reliable, then the second law holds, and so
the entropy conjecture does, i.e., the universe evolves as a Boltzmann process. 
Next, if we apply the norms of Bayesian reasoning, we must condition that
stochastic process on the data we directly have, and nothing else.
In other words, we should fix the Boltzmann process to the only moment about which we directly
have data, namely the present. So we recover the BB hypothesis.
In fact, this is precisely the standard argument for why the BB hypothesis must be true.

In the literature, the BB hypothesis is
commonly formulated this way, and sometimes combined with~\cref{eq:app_B_2_1,eq:app_B_2_2}. 
The resultant reasoning is sometimes referred to as ``unstable''.

If one assumed nothing more than~\cref{eq:app_B_2_1,eq:app_B_2_2} though,
we could only conclude that the probability of the BB hypothesis being true is far greater than
the probability that $\D_1$ is reliable. 
In particular we could {\textit{not}} conclude the value of $B$ is very likely to be the bit, ``not true''.
In other words, this pair of assumptions does not mean the BB hypothesis is likely to be false.

Commonly in the literature people want to go further than just those two assumptions, and assume
that $\D_1$ is highly likely to be reliable.
Doing this though just means we have an instance of~\cref{eq:app_B_2_1,eq:app_B_2_2,eq:app_B_5}. 
Again, we cannot conclude that the value of $B$ must be the bit, ``not true''. Instead
we can only conclude that in point of fact, at least one of our three suppositions for the probabilities must be wrong.

In light of these nuances, the definition of the BB hypothesis in the main text is far more careful. It does not
presume that~\cref{eq:app_B_2_1,eq:app_B_2_2,} hold and / or that~\cref{eq:app_B_5} also holds. 

As mentioned above, it is quite challenging to tease out the physics that
determines which of those three suppositions is wrong.
We can illustrate this issue informally, focusing on the assumption that $P(A = a\,|\,B=b) \simeq 0$.
First note that the entropy conjecture relies on some
very strong assumptions. One of the most glaring is that it assumes that the system in
question is closed to the outside world during its evolution.
This assumption might hold for the case where the system in question is the entire universe 
(though even that assumption is quite problematic). But it certainly does not hold for any memory
system recording the results of experiments we once performed --- \textit{by definition}, such a memory system
is open to the outside world. 

The entropy of such a system can evolve in the opposite direction from that of the universe as a whole,
the entropy conjecture notwithstanding.\footnote{Indeed, the standard thermodynamic understanding of
modern day biological organisms is that they are subsystems of the universe which are tightly coupled to
the outside world, in a special way that causes their entropy to evolve in the opposite direction of the universe
as a whole. Under the BB hypothesis, subsystems whose entropy evolves in the opposite
direction of that of the universe as a whole are equally likely to occur in our past as
in our future. So we could have the entropy of the universe as a whole
increase as one moves further into the past from our present, while the entropy of a subsystem of the universe 
which is coupled to the outside world decreases as one moves further into the past from our present.}
Given this, suppose the BB hypothesis holds, and consider our memory system whose current contents are $\D_1$, which
was coupled to a set of experimental
tests of Newton's laws sometime in our past. Since it has been tightly coupled with the outside world,
the entropy of that memory system might be decreasing into our past even
as the entropy of the entire universe increases into our past. In turn, the decrease of entropy
of the memory system into our past is precisely what one needs for the contents of that memory
to be reliable, according to the standard thermodynamic processes that establish the reliability
of memory~\cite{wolpert2024memory,rovelli2022memory}. In such a scenario, it would \textit{not}
be the case that $P(A = a\,|\,B=b) \simeq 0$.

Stated more prosaically, yes, under the BB hypothesis the present states of our memory systems might just be neurobiological fluctuations,
completely unreliable in that they don't imply what they seem to. But it's also quite possible under the BB hypothesis that
those memory systems of our past are highly reliable, even though (under the BB hypothesis) the entropy of the universe as a
whole is far higher in the past than the present.

\end{document}